\documentclass{ws-mpla}
\usepackage[super]{cite}
\usepackage{graphicx}
\begin{document}

\markboth{S. Jalalzadeh and A. J. S. Capistrano}
{Bohmian mechanics of Klein-Gordon equation via quantum metric and mass}

\catchline{}{}{}{}{}

\title{Bohmian mechanics of Klein-Gordon equation via quantum metric and mass}

\author{ S. Jalalzadeh}

\address{Departamento de F\'isica, Universidade Federal de Pernambuco,\\
Recife, PE, 52171-900, Brazil\\ shahram@df.ufpe.br}

\author{A. J. S. Capistrano}

\address{Federal University of Latin-American Integration,\\
Foz do Igua\c{c}u, Paran\'{a} 85867-970, Brazil\\
Casimiro Montenegro Filho Astronomy Center, Itaipu Technological Park,\\
Foz do Igua\c{c}u, Paran\'{a} 85867-970, Brazil\\
abraao.capistrano@unila.edu.br}

\maketitle


\begin{abstract}
The causal stochastic interpretation of relativistic quantum mechanics has the problems of superluminal velocities, motion backward in time and the incorrect non-relativistic limit. In this paper, according to the original ideas of de Broglie, Bohm and Takabayasi, we have introduced simultaneous quantum mass and quantum metric of curved spacetime to  obtain a correct relativistic theory free of mentioned problems.
\end{abstract}

\ccode{PACS Nos.: 03.65.Ta, 03.65.Ca, 03.65.Pm}

\section{Introduction}
There are various objections to use Klein-Gordon (KG) equation as a particle equation. It is well known the time component of current is not positive definite. Also, the probability density
interpretation of squared modulus of wave function is inconsistent, because its integral through spatial part of spacetime would not be conserved in time. The usual
solution these problems, in flat background Lorentzian manifold, is a rejection of the first
quantized version of the KG equation in
favor of the second quantization formalism given by
Pauli and Weisskopf \cite{Pauli}.
In curved spacetime the situation is more complicated and quantum fields
lose their interpretation as asymptotic particles. Another problem is that unless the manifold has a global timelike killing vector, there is no
way to define vacuum state in a canonical way. Moreover, at the fundamental level,
it is not clear at all why the probability interpretation of wave function in non-relativistic quantum mechanics
is in agreement with experiments, but not in the KG equation \cite{Nik2}. On the other hand, as argued by many authors  \cite{BR},  it seems that the causal stochastic interpretation of quantum mechanics (SIQM) may have a solution
to mentioned problems of relativistic quantum mechanics.
According to the de Broglie-Bohm (dBB) \cite{Db,B1} in SIQM the wave function of quantum system describes a real field, called the quantum
potential, which guides a particle along a trajectory having simultaneously
well-defined position and velocity. The statistical aspects of quantum theory
emerge through the uncontrollable character of both subquanta fluctuations and initial conditions \cite{C1}. The novel feature of this interpretation
is the appearance of the quantum potential, a highly nonlocal function of the parameters of particles and their environments.
In fact, in the causal stochastic interpretation,  it is shown that the claim of the Copenhagen interpretation according to which we must give up the concepts of causality, continuity and the objective reality of individual quantum objects is false \cite{Dur}. On the other hand, the current form of the dBB interpretation of quantum mechanics is not fully satisfactory. The KG relativistic wave function leads to superluminal velocities and motion backward in time \cite{Nik1}. The aim of the present letter is to show that in the SIQM is possible to redefine the quantum mass and quantum metric of  spacetime in which the superluminal velocities and backward motion in time will vanish.

\section{Causal stochastic interpretation of Klein-Gordon equation}
In this section, we assume the background spacetime $({\mathcal M})$ equipped with $(g, \nabla)$
where $g$ is spacetime metric 2-form with signature $(-,+,+,+)$ and $\nabla$
is the Levi-Civita connection. Let $\Phi(x^\mu)$ be the KG wave function
and ${\mathcal A}$ be the external electromagnetic
1-form such as $\mathcal A = A_\mu dx^\mu\in T^*{\mathcal M}$, where $T^*{\mathcal M}$ denotes the dual tangent space. Then the KG equation  for a single particle with mass $m$ and electric
charge $e$ will be \footnote{We use $c=\hbar=1$ throughout this work.}
\begin{eqnarray}\label{1}
\left[g^{\mu\nu}(i\nabla_\mu+eA_\mu)(i\nabla_\nu+eA_\nu)+m^2\right]\Phi(x^\alpha)=0.
\end{eqnarray}
Also, the conserved 4-current, corresponding to the KG equation (\ref{1})
is
\begin{eqnarray}\label{2}
J_\mu=\frac{e}{2im}\left(\Phi^*\nabla_\mu\Phi-\Phi\nabla_\mu\Phi^*\right)-\frac{e^2}{m}A_\mu\Phi\Phi^*.
\end{eqnarray}
The KG equation (\ref{1}) is separable by means of the general complex statement
\begin{eqnarray}\label{3}
\Phi(x^\mu)=R(x^\mu)e^{iS(x^\mu)},
\end{eqnarray}
where $R$ and $S$ are real functions of spacetime coordinates $x^\mu$ $(\mu,\nu=0,1,2,3)$.  Substituting
(\ref{3}) into the KG equation (\ref{1})  gives \cite{2-24-351,2-24-352,2-24-353}
\begin{eqnarray}\label{4}
g^{\mu\nu}\left(\nabla_\mu S-eA_\mu\right)\left(\nabla_\nu S-eA_\nu\right)+m^2\left(1-\frac{\square
R}{m^2R}\right)=0,
\end{eqnarray}
as the real part, and the conserved 4-current
\begin{eqnarray}\label{5}
\nabla_\mu J^\mu=0,
\end{eqnarray}
as the imaginary part, where $\square=g^{\alpha\beta}\nabla_\alpha\nabla_\beta=\frac{1}{\sqrt{-g}}\partial_\mu(\sqrt{-g}g^{\mu\nu}\partial_\nu)$ is the D' Alambert operator. In addition, the 4-current $J_\mu$ can be reduced to
\begin{eqnarray}\label{6}
J_\mu=\frac{e}{m}R^2(\nabla_\mu S-eA_\mu).
\end{eqnarray}
It is worth noting that equation (\ref{4}) looks like the relativistic Hamilton-Jacobi equation (HJ),
modified by dimensionless quantum potential $Q$ such that
\begin{eqnarray}\label{7}
Q=-\frac{1}{m^2}\frac{\square R}{R}.
\end{eqnarray}
The assumption
introduced by the causal approach is that, in an analogous manner to the non-relativistic de Broglie-Bohm quantum mechanics, together with a location
$x^\mu$ a particle has a well-defined 4-momentum
\begin{eqnarray}\label{8}
p^\mu:=\nabla^\mu S= m\left(\frac{dx^\mu}{d\tau} + eA^\mu\right),
\end{eqnarray}
as a set of equations with the aid of which momentum field, or equivalently 4-velocity field
\begin{eqnarray}\label{vel}
u^\mu=\frac{dx^\mu}{d\tau}=\frac{1}{m}(\nabla^\mu S-eA^\mu),
\end{eqnarray}
where $d\tau^2=-g_{\mu\nu}dx^\mu dx^\nu$ denotes the proper time. Inserting (\ref{8}) into the (\ref{5}) gives
$\nabla_\mu(\rho u^\mu)=0$, which is the correct continuity equation for probability
density  (the background fluid density) $\rho=R^2$.  From the modified  HJ equation (\ref{4}), it is clear that the
identity $g_{\mu\nu}u^\mu u^\nu=-1$ is violated if the law of motion (\ref{8})
and (\ref{vel}) can be applied, except in those special cases in which quantum potential vanish
\cite{2-24-351,2-24-352}. Moreover, de Broglie \cite{2-24-351} and Takabayasi \cite{2-24-352} pointed out that this discrepancy
might be corrected, and the law of motion can be maintained by attributing to a particle
a quantum mass $M=m\sqrt{1+Q}$. On the other hand, as pointed out by Takabayasi \cite{2-24-352},
if we use the mentioned quantum mass in equations (\ref{5}), (\ref{6}) and (\ref{8}) we find $\nabla_\mu(MR^2u^\mu)=0$,
then it is no longer possible to consider the imaginary part of KG equation
(\ref{5}) as a continuity equation for background fluid density $\rho$, e.g.
$\nabla_\mu(\rho u^\mu)=0$. Also,
the 4-velocity $u^\mu$ cannot,
in general, have the meaning of a velocity vector field, because the
variable mass $M$ can be imaginary (tachyonic) and the particle trajectory which
start as a timelike trajectory becomes spacelike and can even move backwards in time. An alternative suggestion to solve the incompatibility of the law of
motion with quantum HJ equation and unity of 4-velocity, is to define a new
metric for spacetime which is related conformally to the $g_{\mu\nu}$
\begin{eqnarray}\label{9}
\tilde g_{\mu\nu}=(1+Q)g_{\mu\nu}.
\end{eqnarray}
Then the HJ equation (\ref{4}) will turn to
\begin{eqnarray}\label{10}
\tilde{g}^{\alpha\beta}\left(\tilde\nabla_\mu S-eA_\mu\right)\left(\tilde\nabla_\nu S-eA_\nu\right)+m^2=0,
\end{eqnarray}
where $\tilde\nabla$ denotes the Levi-Civita connection of
$\tilde{g}$.
Inserting $\tilde\nabla^\mu S-eA^\mu=m\frac{dx^\mu}{d\tau}$ into the (\ref{10})
we obtain $\tilde g_{\alpha\beta}u^\alpha u^\beta=-1$. But this is not the
ultimate remedy of the problem. The expectation is that the continuity equation
(\ref{5}), by inserting the above deformed metric and using (\ref{6}), turn to the continuity equation for fluid as $\tilde\nabla_\mu(\rho u^\mu)=0$.
It is easy to see that this is incompatible with the imaginary part of KG
wave equation (\ref{5}).

The last step, deformation of spacetime metric via
quantum potential, sounds interesting in development of relativistic causal
stochastic interpretation of quantum mechanics. As we know, the trajectory of particle
is a hidden variable so as to be able to give a casual interpretation \cite{Tr}.
Hence, it seems that the metric structure of spacetime may be a hidden variable nature too. On the other hand, hidden variable nature of trajectories is in conflict with  localized mass point of view of particles, i.e. the
quantum effects eliminates point-like structures in favor of smeared objects
\cite{Nic}.
Therefore, it may be natural to assume that both  of particle mass and also
the metric of spacetime are dependent to the quantum potential, or precisely
\begin{eqnarray}\label{11}
\begin{array}{cc}
M:=(1+Q)^Am,\\
\hat{g}_{\mu\nu}:=(1+Q)^Bg_{\mu\nu},
\end{array}
\end{eqnarray}
 where $A$ and $B$ are two dimensionless constants and  $\hat\nabla$ denotes the covariant derivative respect to the deformed metric defined in equation (\ref{11}).
Now if we define
the 4-momentum as
\begin{eqnarray}\label{12}
\begin{array}{cc}
p_\mu:=\hat\nabla_\mu S,\\
\hat\nabla_\mu S-eA_\mu=Mu_\mu,
\end{array}
\end{eqnarray}
where $u^\mu=\frac{dx^\mu}{d\tau}$ is the 4-velocity of particle in quantum
spacetime $(\hat{\mathcal M}, \hat g, \hat\nabla)$ and the proper time is
given by $d\tau^2=-\hat{g}_{\mu\nu}dx^\mu dx^\nu$. Then, the consistency of quantum  HJ equation (\ref{4}) and $\tilde{g}_{\mu\nu}u^\mu
u^\nu=-1$ gives $2A+B=1$. Also the continuity equation (\ref{5})
is equivalence to $\nabla_\mu(M\rho u^\mu)=(\sqrt{-g}M\rho)_{,\mu}=0$. To obtain
the correct continuity equation for background fluid with density $\rho$,
i.e.,
$\hat\nabla_\mu(\rho u^\mu)=(\rho\sqrt{-\hat g} u^\mu)_{,\mu}=0$, it
is enough to put $M\sqrt{-g}=m\sqrt{-\hat{g}}$, which gives the second relation $2B=A$.
Hence we obtain
\begin{eqnarray}\label{13}
\begin{array}{cc}
M=(1+Q)^{\frac{2}{5}}m,\\
\hat{g}_{\mu\nu}=(1+Q)^\frac{1}{5}g_{\mu\nu}.
\end{array}
\end{eqnarray}
Inserting equations (\ref{12}) and (\ref{13}) into the (\ref{4}) and (\ref{5})
give the quantum HJ and continuity equations, respectively
\begin{eqnarray}\label{14}
\hat{g}^{\alpha\beta}(p_\alpha-eA_\alpha)(p_\beta-eA_\beta)+M^2=0,
\end{eqnarray}
\begin{eqnarray}\label{15}
\hat\nabla_\mu(\rho u^\mu)=0.
\end{eqnarray}
As we see from equation (\ref{13}), the quantum mass $M$ is real for any
value of quantum potential and consequently the theory is ghost free and the trajectory
of particles remain timelike. The wave function (\ref{3}) is invariant with respect to a change of its phase $S(x^\mu)$ by an integer multiple of $2\pi\hbar$.
Consequently definition of momentum (\ref{12}) gives
\begin{eqnarray}\label{c1}
\oint p_\mu dx^\mu=\oint\left(Mu_\mu+eA_\mu\right)dx^\mu=nh,\hspace{.3cm}n=0,1,2,...\,,
\end{eqnarray}
as a condition of compatibility between equations (\ref{14}) and (\ref{15}) and KG equation (\ref{1}). The path of integration may not pass through points or regions where
probability density is zero because in this kinds of regions according to the equation of continuity (\ref{15}) the velocity field is possibly singular.
To obtain corresponding geodesic equation of quantum particle, we differentiate the 4-momentum defined in (\ref{12}) with respect to the affine parameter of the trajectory,
$d\tau$, which gives $
\frac{dp_\mu}{d\tau}=u^\alpha\partial_\alpha\partial\mu S=u^\alpha\partial_\mu\hat\nabla_\alpha
S$.  Now using (\ref{12}) and (\ref{14}) we find
\begin{eqnarray}\label{16}
\frac{du^\mu}{d\tau}+\hat\Gamma^\mu_{\alpha\beta}u^\alpha u^\beta=-\frac{\partial_\alpha
M}{M}\left(\hat{g}^{\alpha\mu}+u^\alpha u^\mu\right)+\frac{e}{M}F^\mu_{\,\,\,\,\,\alpha}u^\alpha,
\end{eqnarray}
where $F_{\mu\nu}=\partial_\mu A_\nu-\partial_\nu A_\mu$ denotes the electromagnetic field strength tensor. Equations (\ref{14}) and (\ref{15}) are now regarded as an ensemble of particle worldlines which belongs to a single 4-velocity
potential $S(x^\mu)$ by the relation (\ref{12}) and is determined by the classical
relativistic dynamics assuming the modified metric and mass by the quantum
potential that is in turn dependent upon the density distribution of the
ensemble. If a particle traverses a region of space where the Lorentz force,
$(e/M)F^\mu_{\,\,\,\,\nu}u^\nu$, vanishes, but the electromagnetic potentials do not. Then geodesic equation (\ref{16}) explains the Aharonov-Bohm
effect since the particle is at all times acted upon by the quantum force
(first term on the r.h.s. of (\ref{16})) and also moves in deformed space
via quantum metric $\hat{g}$ \cite{G,15,16}. Note that for vanishing values of quantum
conformal factor, $Q+1=1-\frac{\square
R}{m^2R}$, in quantum metric (\ref{13}), the quantum metric
$\hat g$ will be singular. Therefore, the quantum conformal boundary of spacetime
$(\hat{\mathcal M},\hat g,\hat\nabla)$ is given by $\frac{\square
R}{m^2R}=1$. Equivalently, the quantum conformal boundary of spacetime, $\partial\hat{\mathcal
M}$ is
given by those regions of spacetime that $R$ fluctuates greatly of the
order of the Compton wave length of particle, $1/m$. Therefore, the observer
does not have any access to the region outside.

In  the non-relativistic approximation, ($u^0\approx1$, $|dx^i/dt|\ll 1$, $\tilde{g}_{00}\approx-1,i=1,2,3$), the conformal factor reduces to
\begin{eqnarray}\label{17}
Q=-\frac{1}{m^2}\frac{\square R}{R}\approx-\frac{1}{m^2}\frac{\partial_i\partial^i
R}{R}=\frac{2}{m}U,
\end{eqnarray}
where $U=-\frac{1}{2m}\partial_i\partial^iR/R$ is the non-relativistic quantum potential
\cite{B1}. Consequently one can find that the geodesic equation (\ref{16}) reduces to the
Euler equation of the Madelung fluid equation
\begin{eqnarray}\label{18}
\frac{d^2x^i}{dt^2}=-\partial^i(\phi+U)+\frac{e}{m}\left(E^i+\epsilon_{ijk}B^k\frac{dx^j}{dt}\right),
\end{eqnarray}
where $\phi$ is the weak gravitational potential, $E^i$ and $B^i$ are the components of electric and magnetic fields, respectively. As expected, we obtain the
correct non-relativistic limit \cite{B1} and the particle feels the usual Lorentz, weak gravitational field plus the non-relativistic quantum forces.

\section{Conclusion}
It was shown that with simultaneous quantum conformal transformations of mass
of relativistic particles and the metric of spacetime, it is possible to write a covariant casual stochastic quantum equations of motion for a relativistic
particle with no problems of superluminal velocities and the incorrect non-relativistic limit. This
shows that the inertia of a quantum particle and metric structure of spacetime
are determined by a scalar field which is a function of all relevant quantum processes occurring in the environment of the particle. Therefore, the quantum
potential plays the role of the scalar field introduced by Brance and
Dicke to supplement Einstein's general theory of relativity with an explicit statement of Mach's principle.


\end{document}